\documentclass[12pt]{article}
\usepackage[dvips]{color}
\usepackage{epsfig}
\usepackage{amsmath}
\usepackage[colorlinks]{hyperref}
\usepackage{graphicx}

\begin{document}
\title{Interacting Entropy-Corrected Holographic Dark Energy and IR
Cut-Off Length}
\author{J. Sadeghi$^{a}$\thanks{Email: pouriya@ipm.ir},\hspace{1mm}
B. Pourhassan$^{b}$\thanks{Email:
b.pourhassan@umz.ac.ir}\hspace{1mm}
and Z. Abbaspour Moghaddam$^{a}$\thanks{Email: kmabbaspour@yahoo.com}\\
$^a$ {\small {\em Faculty of Basic  Sciences, Departments of Physics, Mazandaran University,}}\\
{\small {\em P.O.Box 47416-95447, Babolsar, Iran}}\\
$^b$ {\small {\em  Department of Physics, I.H.U., Tehran, Iran}}}
\maketitle
\begin{abstract}
In this paper we consider holographic dark energy model with
corrected holographic energy density and show that this model may be
equivalent to the modified Chaplygin gas model. Then we obtain
relation between entropy corrected holographic dark energy model and
scalar field models. We do these works by using choices of IR
cut-off length proportional to the Hubble radius, the event horizon
radius, the Ricci length, and the Granda-Oliveros
length.\\\\
\noindent {\bf Keywords:} Holographic Dark Energy; Modified
Chaplygin Gas; Scalar Field.
\end{abstract}
\section{Introduction}
Astrophysical observations of type Ia supernovae [1, 2], the
large-scale structure [3], the cosmic microwave back grand radiation
[4] suggest accelerated expansion of universe which may be described
by the dark energy [5, 6]. The simplest model to describe dark
energy is the cosmological constant, but this model suffers from the
cosmic coincidence problems and the fine-tuning. Another suggested
models consist of k-essence [7, 8], tachyon [9], quintom [10],
quintessence [11] and phantom [12]. Also, there is an interesting
model based on equation of state of Chaplygin gas and its
generalizations [13-20]. On the other hand, one can investigate the
nature of dark energy based on some principles of quantum gravity,
which is yield to holographic dark energy (HDE) model [21-29].\\
In the recent work [30], interacting HDE based on scalar field
models has been studied and found that interacting HDE model can be
considered as the modified Chaplygin gas model. In that case it is
possible to study interacting entropy-corrected holographic dark
energy models [31, 32, 33]. So, in the Refs. [31, 32] a
correspondence between the tachyon, K-essence and dilaton scalar
field models established with the interacting entropy-corrected
holographic dark energy model in non-flat FRW universe, and dynamics
of these scalar fields according to the evolutionary behavior of the
interacting entropy-corrected holographic dark energy model has been
studied. Then Ref. [33] extended previous works to the case of
variable gravitational constant and obtained the equation of state
and the deceleration parameters of the interacting viscous
entropy-corrected holographic dark energy model. Also, the potential
and the dynamics of the quintessence, tachyon, K-essence and dilaton
scalar field models according to the evolutionary behavior of the
interacting viscous entropy-corrected holographic dark energy model
with time-varying $G$ reconstructed.\\
Now, in this paper we would
like to obtain relation between interacting entropy-corrected HDE
model and modified Chaplygin gas, also with scalar fields by using
different choices of IR cut-off length.

\section{Entropy-corrected holographic dark energy}
It is found that the HDE density may be obtained from the entropy
bound [34]. In the framework of black hole thermodynamics [35-46],
there is a maximum entropy which is called the Bekenstein-Hawking
entropy bound ($S_{BH}=A/4G$), which is yield to the following
relation for the HDE density [33],
\begin{equation}\label{1}
\rho_d\leq3c^2M^2_pL^{-2},
\end{equation}
where $c$ is a numerical constant (however it may be considered as a
non-constant parameter [47, 48]), $M$ is the reduce Planck mass, and
$L$ shows the IR cut-off length. In the restriction (1), equality is
given if the holographic bound is saturated. In order to consider
the quantum effects the Bekenstein-Hawking entropy may be corrected
as the following [49, 50, 51],
\begin{equation}\label{2}
S= \frac {A}{4G}+\widetilde{\gamma} \ln(\frac
{A}{4G})+\widetilde{\beta},
\end{equation}
where $\widetilde{\gamma}$ and $\widetilde{\beta}$ are undetermined
constants. By using the relation (2) the entropy-corrected HDE
proposed in the following form [52],
\begin{equation}\label{3}
\rho_d=3c^2L^{-2}+\gamma{L^{-4}}\ln{L^2}+\beta{L^{-4}},
\end{equation}
where $\gamma$ and $\beta$ are dimensionless constants. In this
paper we use several choice of IR cut-off length such as $L=H^{-1}$
(the Hubble radius), $L=R_E$ (the radius of the event horizon),
$L=(H^2+\dot{H})^\frac{-1}{2}$ (the Ricci length), and
$L=(\alpha{H}^2+\beta\dot{H} )^\frac{-1}{2}$ (the Granda-Oliveros
length).
\section{Entropy corrected HDE model and modified Chaplygin gas}
As we know the FRW universe is described by the following line
element,
\begin{equation}\label{4}
ds^2=-dt^2+a^2(t)\left(\frac{dr^2}{1-kr^2}+r^2d\Omega^2\right),
\end{equation}
where $k=0, +1, -1$ are corresponding to flat, closed and open
universe. The first Friedman equation is given by,
\begin{equation}\label{5}
H^2+\frac{k}{a^2}=\frac{8\pi G}{3}(\rho_d+\rho_m),
\end{equation}
where $\rho_d$ is the energy density for the entropy corrected
single HDE model and $\rho_m$ is the energy density of dark matter.
Also the conservation energy equation separated as the following,
\begin{equation}\label{6}
\dot{\rho_m}+3H\rho_m=Q,
\end{equation}
and
\begin{equation}\label{7}
\dot{\rho_d}+3H\rho_d(1+\omega)=-Q,
\end{equation}
where equation of state $\omega=\frac{P_d}{\rho_d}$ is used. Also,
$Q$ is the interacting term of the form,
\begin{equation}\label{8}
Q=3H\lambda\rho_{d},
\end{equation}
where $\lambda$ is a dimensionless positive parameter. Therefore we
can rewrite equations (6) and (7) as the following,
\begin{equation}\label{9}
\dot{\rho_m}+3H\rho_m(1-\frac{\lambda}{u})=0,
\end{equation}
and
\begin{equation}\label{10}
\dot{\rho_d}+3H\rho_d(1+\lambda+\omega)=0,
\end{equation}
where $u=\frac{\rho_m}{\rho_d}$. Adding both equations (6) and (7)
gives,
\begin{equation}\label{11}
\dot{\rho_t}+3H(1+\omega_t)\rho_t=0,
\end{equation}
where ${\rho_t=\rho_d+\rho_m}$, and
\begin{equation}\label{12}
\omega_t=\frac{P_d}{\rho_{d}+\rho_{m}}=\frac{\omega\Omega_d}{1+\Omega_k},
\end{equation}
where the fractional energy densities are defined as,
\begin{equation}\label{13}
\Omega_k=\frac{k}{a^2H^2},
\end{equation}
\begin{equation}\label{14}
\Omega_d=\frac{\rho_d}{3M^2_pH^2},
\end{equation}
\begin{equation}\label{15}
\Omega_m=\frac{\rho_m}{3M^2_pH^2},
\end{equation}
which yield to the following relation,
\begin{equation}\label{16}
1+\Omega_k=\Omega_d+\Omega_m.
\end{equation}
In the following we use unitary transformation $M^2_p=8\pi G=1$.\\
Now we investigate relation between entropy corrected HDE model and
modified Chaplygin gas with the following equation of state,
\begin{equation}\label{17}
P=A\rho-\frac{B}{\rho^n}
\end{equation}
with $0<A\leq1$, $B>0$, and $n>0$. We would like to examine the
equation (3) for the following cases.

\subsection{Hubble radius as IR cut-off}
In that case we use,
\begin{equation}\label{18}
L=H^{-1},
\end{equation}
in the equation (3) which gives,
\begin{equation}\label{19}
\rho_d=3c^{2}H^{-2}+\gamma{H^{-4}}\ln{H^{2}}+\beta{H^{-4}}.
\end{equation}
Using the equations (5), (10) and (19) give the following equation
of state parameter,
\begin{equation}\label{20}
\omega=-1+\frac {uC-\lambda}{1-C},
\end{equation}
where $C\equiv c^2-\frac{H^2}{3}(4\gamma\ln{H}+\gamma+2\beta)$. So,
one can obtain,
\begin{equation}\label{21}
\omega_t=(-1+\frac{uC-\lambda}{1-C})\Omega_d.
\end{equation}
On the other hand, the MCG model has the following equation of state
parameter,
\begin{equation}\label{22}
\omega_{CG}=A-\frac{B}{\rho^{n+1}},
\end{equation}
where we used the equation (17) and $\omega_{CG}=P/\rho$. By using
modified Chaplygin gas equation of state in the energy conservation
relation one can obtain scale factor-dependent energy density as the
following,
\begin{equation}\label{23}
\rho^{n+1}=\frac{1}{A+1}\left(B+\frac{\delta}{a^\mu}\right),
\end{equation}
where $\mu=3(A+1)(n+1)$ and $\delta$ is the integration constant.
Substituting (23) in (22) and assuming $\omega_{CG}=\omega_{t}$
give,
\begin{equation}\label{24}
A-\frac{B(A+1)}{B+\frac{\delta}{a^\mu}}=(-1+\frac{uC-\lambda}{1-C})\Omega_{d}.
\end{equation}
Therefore we obtain,
\begin{equation}\label{25}
\sqrt\frac{\rho_{d}(0)}{3}(t-t_0)=\int{\frac{{a}^{\frac{1}{2}(3C(u-\lambda)-2)}}{\sqrt{1-\frac{uC-\lambda}{1-C}}}
{\sqrt{\frac{B-\frac{\delta A}{a^\mu}}{B+\frac{\delta}{a^\mu}}}}}da
\end{equation}
Thus, the entropy-corrected HDE model for Hubble radius as IR
cut-off can be considered as modified Chaplygin gas model.

\subsection{Event horizon radius as IR cut-off}
In that case we use,
\begin{equation}\label{26}
L=R_E,
\end{equation}
in the equation (3). For the event horizon, the radius is chosen in
the form of $L=ar(t)$ where the function $r(t)$ obtained from the
following relation,
\begin{equation}\label{27}
\int{\frac{dr}{\sqrt{1-kr^2}}}=\int{\frac{dt}{a}}={\frac{R_E}{a}},
\end{equation}
with $r(t)=\frac{1}{\sqrt{k}}\sin{y}$. Therefore we can obtain,
\begin{equation}\label{28}
\omega=-1-\lambda-{\frac{1}{\sqrt{3}}}\mathcal{A}+\sqrt{\frac{\Omega_d}{3}}D,
\end{equation}
where
\begin{eqnarray}\label{29}
\mathcal{A}&\equiv&c^2+{\frac{L^{-2}}{3}}(\gamma\ln{L^2}+\beta)D,\nonumber\\
D&\equiv&\frac{-6c^2-8\gamma(L^{-2}\ln{L}-1)-4\beta{L^{-2}}}{(3c^2+L^{-2}(\gamma\ln{L}+\beta))^{\frac{3}{2}}}.
\end{eqnarray}
Now from the equation (12) we have,
\begin{equation}\label{30}
-\frac{B-\frac{\delta
A}{a^\mu}}{B+\frac{\delta}{a^\mu}}=(-1-\lambda-\frac{1}{\sqrt{3}}\mathcal{A}+\sqrt{\frac{\Omega_d}{3}}D)\Omega_{d},
\end{equation}
which may be rewritten as the following equation,
\begin{equation}\label{31}
\theta{x^3}-\tau{x^2}+\xi=0
\end{equation}
where $\theta=\frac{D}{\sqrt{3}}$,
$\tau=(1+\lambda+\frac{\mathcal{A}}{\sqrt{3}})$,
$\xi=\frac{B-\frac{\delta A}{a^\mu}}{B+\frac{\delta A}{a^\mu}}$, and
$x=\sqrt{\Omega_d}$. Hence, we obtain,
\begin{eqnarray}\label{32}
\sqrt{\Omega_d}&=&{\frac {\sqrt
[3]{-108\,\xi\,{\theta}^{2}+8\,{\tau}^{3}+12\, \sqrt {-3\,\xi\,
\left( -27\,\xi\,{\theta}^{2}+4\,{\tau}^{3} \right) }
\theta}}{6\theta}}\nonumber\\
&+&{\frac {2{\tau}^{2}}{3\theta\,\sqrt [3]{-108\,\xi
\,{\theta}^{2}+8\,{\tau}^{3}+12\,\sqrt {-3\,\xi\, \left( -27\,\xi\,{
\theta}^{2}+4\,{\tau}^{3} \right) }\theta}}}+{\frac {\tau}{3\theta
}}.
\end{eqnarray}
From the Fig. 1 we can see that the parameter $\gamma$ increased the
$\Omega_{d}$ while the parameter $\beta$ decreased one.

\begin{figure}[th]
\begin{center}
\includegraphics[scale=.3]{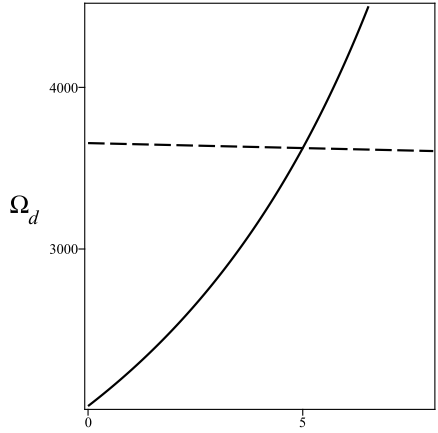}
\caption{Typical behavior of $\Omega_{d}$ with $\gamma$ (solid line)
and $\beta$ (dashed line).}
\end{center}
\end{figure}
\subsection{Ricci length scale as IR cut-off}
In that case we use,
\begin{equation}\label{33}
L=(2H^2+\dot{H})^\frac{-1}{2},
\end{equation}
in the equation (3), and after some calculations obtain,
\begin{equation}\label{34}
\Omega_d=\frac{7B-\delta(\frac{3A-4}{a^\mu})}{(B+\frac{\delta}{a^\mu})(2\eta+3(u+1))},
\end{equation}
where, $\eta=c^{2}+\frac{1}{3L^{2}}(\gamma\ln{L}+\beta)$. We find
that $\Omega_d$ is always positive (see Fig. 2), which is agree with
the Ref. [30]

\begin{figure}[th]
\begin{center}
\includegraphics[scale=.3]{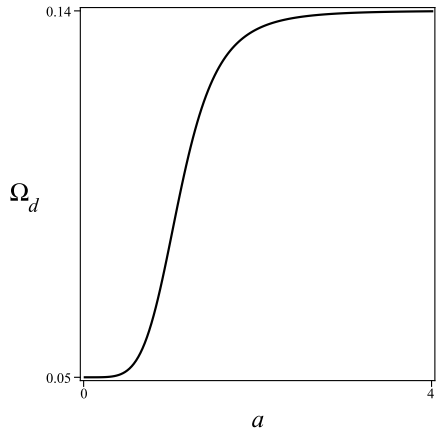}
\caption{Typical behavior of $\Omega_{d}$ in terms of scale factor.}
\end{center}
\end{figure}
\subsection{Granda-oliveros length scale as IR cut-off length}
In that case we use [53, 54],
\begin{equation}\label{35}
L=(\alpha{H^2}+\beta\dot{H})^\frac{-1}{2},
\end{equation}
in the equation (3), and after some calculations and compare with
modified Chaplygin gas obtain,
\begin{equation}\label{36}
\Omega_d=\frac{B(3\beta+2\alpha)+\delta(\frac{-3\beta
A+2\alpha}{a^\mu})}{3\beta(B+\frac{\delta}{a^\mu})(\frac{2}{3\beta}+(u+1))}.
\end{equation}
\section{Entropy-corrected HDE model and scalar field models}
In this section we consider three models of scalar field and
investigate relation with entropy corrected HDE model.

\subsection{K-essence}
The general form of the action for K-essence scalar field written as
the following,
\begin{equation}\label{37}
S=\int{d^4{X}\sqrt{-g}P(\Phi,X)},
\end{equation}
where $P(\Phi,X)$ is pressure density and
$X=\frac{-{\dot{\Phi}}^2}{2}$. The energy density and pressure
represented as the following,
\begin{equation}\label{38}
\rho(\Phi,X)=f(\Phi)(-X+3X^2),
\end{equation}
and
\begin{equation}\label{39}
P(\Phi,X)=f(\Phi)(-X+X^2),
\end{equation}
where $f(\Phi)$ represents the K-essence potential, and in
comparison with entropy corrected HDE model we obtain,
\begin{equation}\label{40}
X=\frac{1-\omega}{1-3\omega}.
\end{equation}
\subsubsection{Event horizon radius as IR cut-off}
In that case by using the relation (26) we can obtain,
\begin{equation}\label{41}
X=\frac{2+\lambda+\frac{\mathcal{A}}{\sqrt{3}}-\sqrt{\frac{\Omega_d}{3}}D}{4+3\lambda+\sqrt{3}\mathcal{A}-\sqrt{3\Omega_d}D},
\end{equation}
which yields to the following expression,
\begin{equation}\label{42}
f(\Phi)=\frac{\rho_d}{3X^2-X}=\frac{3H^2\Omega_d(4+3\lambda+\sqrt{3}\mathcal{A}-\sqrt{3\Omega_d}D)}{2(2+\lambda+\frac{\mathcal{A}}{\sqrt{3}}-\sqrt{\frac{\Omega_d}{3}}D)}.
\end{equation}
In order to have a real K-essence scalar field we should have the
following restriction,
\begin{equation}\label{43}
\frac{1}{3}(\frac{4+3\lambda+\sqrt{3}\mathcal{A}}{D})^2<\Omega_d<3(\frac{2+\lambda+\frac{\mathcal{A}}{\sqrt{3}}}{D})^{2}.
\end{equation}
\subsubsection{Ricci length scale as IR cut off}
Now, by using the relation (33) we obtain,
\begin{equation}\label{44}
X=\frac{\frac{2}{3\Omega_d}[\frac{\Omega_d}{\eta}-2]+(u+2)}{\frac{2}{\Omega_d}[\frac{\Omega_d}{\eta}-2]+(3u+4)}.
\end{equation}
Therefore, we obtain,,
\begin{equation}\label{45}
f(\Phi)=\frac{3H^2\Omega_d[\frac{2}{\Omega_d}(\frac{\Omega_d}{\eta}-2)+(3u+4)]^2}{2(2+\frac{2}{3\Omega_d}(\frac{\Omega_d}{\eta}-2)+u)}.
\end{equation}
In order to have a real K-essence scalar field we should have the
following restriction,
\begin{equation}\label{46}
\frac{4}{\frac{2}{\eta}+3u+6}<\Omega_d<\frac{4}{\frac{2}{\eta}+3u+4}.
\end{equation}
From the equation,
\begin{equation}\label{47}
{\dot{\Phi}}^2=-2X,
\end{equation}
we can write the explicit form of the K-essence scalar field as,
\begin{equation}\label{48}
\Phi=\sqrt{2}\int{\sqrt{\frac{\frac{2}{3\Omega_d}(\frac{\Omega_{d}}{\eta}-2)+(u+2)}{\frac{-2}{\Omega_d}(\frac{\Omega_d}{\eta}-2)-(3u+4)}}}dt.
\end{equation}
\subsection{Tachyon field}
The effective lagrangian for the tachyon field is described by,
\begin{equation}\label{49}
L=-V(\Phi)\sqrt{1-g^{\mu\nu}{\partial\mu}^{\Phi}{\partial\nu}^\Phi}.
\end{equation}
The energy density and pressure for the tachyon field are given by
the following expressions,
\begin{equation}\label{50}
\rho=\frac{V(\Phi)}{\sqrt{1-{\dot{\Phi}}^2}},
\end{equation}
and,
\begin{equation}\label{51}
P=-V(\Phi)\sqrt{1-{\dot{\Phi}}^2}.
\end{equation}
In order to find relation between tachyon field and entropy
corrected HDE model we consider cases of event horizon radius and
Ricci length scale as IR cut-off.
\subsubsection{Event horizon radius as IR cut-off}
In that case by using the relation (26) we can obtain,
\begin{equation}\label{52}
\Phi=\int{\frac{\sqrt{-\lambda-{\frac{\mathcal{A}}{\sqrt{3}}+\sqrt{\frac{\Omega_d}{3}}D}}}
{H\Omega_d[-3+(1-\Omega_d)(\sqrt{3}\mathcal{A}-\sqrt{3}\Omega_{d}D)-3(1+\lambda)\Omega_d]}}d\Omega_d+\Phi_{0},
\end{equation}
and,
\begin{equation}\label{53}
V(\Phi)=3\sqrt{1+\lambda+\frac{\mathcal{A}}{\sqrt{3}}-\sqrt{\frac{\Omega_d}{3}}D}H^{2}\Omega_{d}.
\end{equation}
\subsubsection{Ricci length scale as IR cut off}
Now, by using the relation (33) we obtain,
\begin{equation}\label{54}
\Phi=\int{\sqrt{2-\frac{2}{3\Omega_d}(\frac{\Omega_d}{\eta}-2)-u}}dt+\Phi_{0}
\end{equation}
and,
\begin{equation}\label{55}
V(\Phi)=3\sqrt{\frac{2}{3\Omega_d}(\frac{\Omega_d}{\eta}-2)+(u-1)}H^{2}\Omega_d.
\end{equation}
Relation (55) tells that the correction parameters which involves in
$\eta$ decreased potential $V(\Phi)$.
\subsection{Quintessence}
The energy density and pressure for the quintessence are given by,
\begin{equation}\label{56}
\rho_\Phi=\frac{1}{2}{\dot{\Phi}}^2+V(\Phi),
\end{equation}
and
\begin{equation}\label{57}
P_\Phi=\frac{1}{2}{\dot{\Phi}}^2-V(\Phi),
\end{equation}
where $\Phi$ is a quintessence field with potential $V(\Phi)$. Here,
we assumed $\rho_\Phi=\rho_d$ and $P\Phi=P_d$. Therefore, one can be
obtain,
\begin{equation}\label{58}
\Phi=\int{\frac{\sqrt{(1+\omega_d)}\sqrt{(3\Omega_d)}}{\Omega_{d}^{\prime}}}d\Omega_{d},
\end{equation}
and,
\begin{equation}\label{59}
V(\Phi)=\frac{3}{2}(1-\omega_d)H^2\Omega_{d}
\end{equation}
where,
\begin{equation}\label{60}
\Omega_{d}^{\prime}=\Omega_d[-3+(1-\Omega_d)(\sqrt{3}\mathcal{A}-\sqrt{3\Omega_d}D)-3(1+\lambda)\Omega_d]
\end{equation}
\subsubsection{Event horizon radius as IR cut-off}
In this case one obtain,
\begin{equation}\label{61}
\Phi=2\int{\frac{\sqrt{\frac{1}{\sqrt{3}}(xD-\mathcal{A})-\lambda}}{[-\sqrt{3}+(1-x^2)(\mathcal{A}-xD)-\sqrt{3}(1+\lambda)x^2]}}dx+\Phi_0
\end{equation}
where $x\equiv\sqrt{\Omega_d}$ defined as before, and,
\begin{equation}\label{62}
V(\Phi)=\frac{3}{2}H^2(2+\lambda+\frac{\mathcal{A}}{\sqrt{3}}-\sqrt{\frac{\Omega_d}{3}}D)\Omega_{d}.
\end{equation}
Thus, $\Phi$ and $V(\Phi)$ are obtain as a function of the density
parameter $\Omega_d$.
\subsubsection{Ricci length scale as IR cut-off}
By using the relation (33) we obtain,
\begin{equation}\label{63}
\Phi=\sqrt{4-[\frac{2}{\eta}-3u][\frac{7Ba^\mu+\delta(3A+4)}{(Ba^\mu+\delta)(2\eta+3(u+1))}]}\frac{da}{a}+\Phi_{0},
\end{equation}
and,
\begin{equation}\label{64}
V(\Phi)=(3+\frac{1}{\eta}+\frac{3}{2}u)(\frac{7Ba^\mu-\delta(3A-4)}{(Ba^\mu+\delta)(2\eta+3(u+1))}).
\end{equation}
It is clear that the entropy correction parameters decreased value
of potential.
\section{Conclusion}
In this paper, entropy-corrected holographic dark energy model
considered for various IR cut-off such as Hubble radius, event
horizon radius, Ricci length scale, and Granda-Oliveros length
scale. Indeed we extended the Ref. [30] to the case of entropy
corrected with parameters $\gamma$ and $\beta$. We investigated
relation of entropy-corrected holographic dark energy model with
modified Chaplygin gas and scalar field models. We found that
correction parameters may be increased or decreased dark energy
density. Indeed, we found that the first correction term (second
terms of right hand side of the equation (3)) increased dark energy
density, while the second correction includes parameter $\beta$
decreased dark energy density. Therefore appropriate choices of
$\gamma$ and $\beta$ may be canceled each other. Also we found that
the correction parameters
reduced potential of scalar fields.\\
Here, there are some interesting problems such as adding shear and
bulk viscosity to system and repeat discussion of this paper. Also
one can use explicit expression of scale factor in terms of time and
discuss time-dependent solutions.


\begin{thebibliography}{99}
\bibitem{P1}
A.G. Riess et al., Astron. J. 116 (1998) 1009
\bibitem{P2}
S. Perlmutter et al., Astrophys. J. 517 (1999) 565
\bibitem{P3}
M. Tegmark et al., Phys. Rev. D69 (2004) 103501
\bibitem{P4}
D. N. Spergel et al., Astrophys. J. Suppl. Ser.148 (2003) 175
\bibitem{P5}
A.K. Yadav, F. Rahaman, S. Ray, International Journal of Theoretical
Physics 50 (2011) 871
\bibitem{P6}
H. Saadat, International Journal of Theoretical Physics 50 (2011)
140
\bibitem{P7}
C. Armendariz-Picon, V. F. Mukhanov, and P. J. Steinhardt, Phys.
Rev. Lett. 85 (2000) 4438
\bibitem{P8}
C. Armendariz-Picon, V. F. Mukhanov, and P. J. Steinhardt, Phys.
Rev. D 63 (2001) 103510
\bibitem{P9}
M. R. Setare, J. Sadeghi, A.R. Amani, Phys. Lett. B673 (2009) 241
\bibitem{P10}
B. Feng, X.L. Wang, and X.M. Zhang, Phys. Lett. B 607 (2005) 35
\bibitem{P11}
C. Wetterich, Nucl. Phys. B 302 (1988) 668
\bibitem{P12}
R.R. Caldwell, Phys. Lett. B 545 (2002) 23
\bibitem{P13}
H. Saadat and  B. Pourhassan "Viscous Varying Generalized Chaplygin
Gas with Cosmological Constant and Space Curvature", International
Journal of Theoretical Physics, (2013) DOI10.1007/s10773-013-1676-2
\bibitem{P14}
B. Pourhassan, International Journal of Modern Physics D Vol. 22,
No. 9 (2013) 1350061 [arXiv:1301.2788 [gr-qc]]
\bibitem{P15}
A.R. Amani and  B. Pourhassan, International Journal of Theoretical
Physics 52 (2013) 1309
\bibitem{P16}
H. Saadat and  B. Pourhassan, Astrophysics and Space Science 344
(2013) 237
\bibitem{P17}
H. Saadat and  B. Pourhassan, Astrophysics and Space Science 343
(2013) 783
\bibitem{P18}
H. Saadat, H. Farahani, International Journal of Theoretical Physics
52 (2013) 1160
\bibitem{P19}
J. Sadeghi, H. Farahani, "Interaction between viscous varying
modified cosmic Chaplygin gas and Tachyonic fluid", Astrophysics and
space science journal (2013), [arXiv:1304.6987 [gr-qc]]
\bibitem{P20}
H. Saadat, B. Pourhassan, "Effect of Varying Bulk Viscosity on
Generalized Chaplygin Gas", [arXiv:1305.6054 [gr-qc]]
\bibitem{P21}
M.R. Setare, Chin. Phys. Lett. 26 (2009) 029501
\bibitem{P22}
S.D.H. Hsu, Phys. Lett. B 594 (2004) 13
\bibitem{P23}
M.R. Setare, E. N. Saridakis, Phys. Lett. B671 (2009) 331
\bibitem{P24}
M. Jamil, E.N. Saridakis, M.R. Setare, Phys. Lett. B679 (2009) 172
\bibitem{P25}
J. Lu, E.N. Saridakis, M.R. Setare, L. Xu, JCAP 1003 (2010) 031
\bibitem{P26}
M.R. Setare, M. Jamil, JCAP 1002 (2010) 010
\bibitem{P27}
M.R. Setare, M. Jamil, Phys. Lett. B690 (2010) 1
\bibitem{P28}
Kh. Saaidi, A. Aghamohammadi, M.R. Setare, Astrophys. Space Sci. 332
(2011) 503
\bibitem{P29}
M.R. Setare, Elias C. Vagenas, Phys. Lett. B666 (2008) 111
\bibitem{P30}
S. Pan, S. Chakraborty, "Interacting Holographic Dark Energy: Scalar
Field Models", [arXiv:1210.0396 [gr-qc]]
\bibitem{P31}
A. Khodam-Mohammadi, M. Malekjani, Commun. Theor. Phys. 55 (2011)
942
\bibitem{P32}
K. Karami, M.S. Khaledian, M. Jamil, Phys. Scr. 83 (2011) 025901
\bibitem{P33}
F. Adabi, K. Karami, F. Felegary, Z. Azarmi, Res. Astron. Astrophys.
12 (2012) 26
\bibitem{P34}
B. Guberina, R. Horvat, H. Nikolic, JCAP 01 (2007) 012
\bibitem{P35}
A. Pourdarvish, J. Sadeghi, H. Farahani, and B.Pourhassan,
International Journal of Theoretical Physics, DOI:
10.1007/s10773-013-1658-4
\bibitem{P36}
J.D. Bekenstein, Phys. Rev. D 9 (1974) 3292
\bibitem{P37}
J. Sadeghi , B. Pourhassan  and  F. Pourasadollah, Physics Letters B
720 (2013) 244 [arXiv:1209.1874 [hep-th]]
\bibitem{P38}
J.D. Bekenstein, Phys. Rev. D 7 (1973) 2333
\bibitem{P39}
J. Sadeghi, K. Jafarzade, and B. Pourhassan, "Thermodynamical
Quantities of Horava-Lifshitz Black Hole", International Journal of
Theoretical Physics 51 (2012) 3891
\bibitem{P40}
J.D. Bekenstein, Phys. Rev. D 23 (1981) 287
\bibitem{P41}
J. Sadeghi, M. R Setare and B. Pourhassan, Acta Physica Polonica B
40 (2)  (2009) 251 [arXiv:0707.0420 [hep-th]]
\bibitem{P42}
J.D. Bekenstein, Phys. Rev. D 49 (1994) 1912
\bibitem{P43}
J. Sadeghi,  M.R. Setare,  B. Pourhassan,  Eur. Phys. J. C 53 (2008)
95
\bibitem{P44}
S.W. Hawking, Commun. Math. Phys. 43 (1975) 199
\bibitem{P45}
J. Sadeghi, B. Pourhassan, A. Asadi, "Thermodynamics of string black
hole with hyperscaling violation", [arXiv:1209.1235 [hep-th]]
\bibitem{P46}
S.W. Hawking, Phys. Rev. D 13 (1976) 191
\bibitem{P47}
H. Saadat, International Journal of Theoretical Physics 52 (2013)
1027
\bibitem{P48}
N. Radicella, D. Pavon, J. Cosmol. Astropart. Phys. 10 (2010) 005
\bibitem{P49}
R. Banerjee, B.R. Majhi, Phys. Lett. B 662 (2008) 62
\bibitem{P50}
R. Banerjee, B.R. Majhi, JHEP 06 (2008) 095
\bibitem{P51}
J. Zhang, Phys. Lett. B 668 (2008) 353
\bibitem{P52}
H. Wei, Commun. Theor. Phys. 52 (2009) 743
\bibitem{P53}
L.N. Granda, A. Oliveros, Phys. Lett. B669 (2008) 275
\bibitem{P54}
L.N. Granda, A. Oliveros, Phys. Lett. B671 (2009) 199
\end{thebibliography}
\end{document}